\documentclass[12pt,aps,superscriptaddress,prl]{revtex4}
\usepackage{amsmath}
\usepackage{amsthm}
\usepackage[colorlinks=true,urlcolor=blue,citecolor=blue,linkcolor=black]{hyperref}
\usepackage[shortlabels]{enumitem}
\usepackage{amsfonts}
\usepackage{graphicx}
\usepackage{bbold}
\usepackage{stmaryrd}
\usepackage{color}
\usepackage{hyperref}
\usepackage{verbatim}
\usepackage{cases}
\usepackage{amsfonts}
\usepackage{amssymb}
\usepackage[]{mathrsfs}
\usepackage{xcolor}
\usepackage{anysize}
\usepackage{epsfig}
\usepackage{bm}
\usepackage{setspace}
\usepackage{enumerate}
\usepackage{dcolumn}% Align table columns on decimal point
\usepackage{bm}% bold math
\usepackage{enumitem}
\newcommand{\bra}[1]{\left< #1 \right\vert}
\newcommand{\ket}[1]{\left\vert #1 \right>}

\newcommand{\pare}[1]{\left( #1 \right)}

\newcommand{\im}{\mathrm{i}}
\newcommand{\p}{\psi}
\newcommand{\PP}{\Psi}
\newcommand{\g}{\gamma}
\newcommand{\cc}{\kappa}
\newcommand{\ep}{\epsilon}

\renewcommand{\epsilon}{\varepsilon}

\begin{document}

\title{Endurance of Quantum Coherence \\in Born-Markov Open Quantum Systems}

\date{\today}

\author{Armando Perez-Leija}\email{apleija@gmail.com}
\affiliation{Max-Born-Institut, Max-Born-Strasse 2A, 12489 Berlin, Germany}
\affiliation{Humboldt-Universit\"at zu Berlin, Institut f\"{u}r Physik, AG Theoretische Optik \& Photonik, Newtonstrasse 15, 12489 Berlin, Germany}
\affiliation{Institute of Applied Physics, Friedrich-Schiller Universit\"at, Max-Wien-Platz 1, 07743 Jena, Germany}
\author{Diego Guzm\'{a}n-Silva}
\affiliation{Institute of Applied Physics, Friedrich-Schiller Universit\"at, Max-Wien-Platz 1, 07743 Jena, Germany}
\author{Roberto~de~J.~Le\'{o}n-Montiel}\email{roberto.leon@nucleares.unam.mx}
\affiliation{Instituto de Ciencias Nucleares, Universidad Nacional Aut\'onoma de M\'exico, 70-543, 04510 Cd. Mx., M\'exico}
\author{Markus Gr\"{a}fe}
\affiliation{Institute of Applied Physics, Friedrich-Schiller Universit\"at, Max-Wien-Platz 1, 07743 Jena, Germany}
\author{Matthias Heinrich}
\affiliation{Institute of Applied Physics, Friedrich-Schiller Universit\"at, Max-Wien-Platz 1, 07743 Jena, Germany}
\author{Hector Moya-Cessa}
\affiliation{Instituto Nacional de Astrof\'{i}sica, \'{O}ptica y Electr\'{o}nica, Calle Luis Enrique Erro 1, Santa Mar\'{i}a
Tonantzintla, Puebla CP 72840, M\'{e}xico}
\author{Kurt Busch}
\affiliation{Humboldt-Universit\"at zu Berlin, Institut f\"{u}r Physik, AG Theoretische Optik \& Photonik, Newtonstrasse 15, 12489 Berlin, Germany}
\affiliation{Max-Born-Institut, Max-Born-Strasse 2A, 12489 Berlin, Germany}
\author{Alexander Szameit}
\affiliation{Institut f\"ur Physik, Universit\"at Rostock, D-18051 Rostock, Germany}

\begin{abstract}
\begin{center}
\textbf{ABSTRACT} 
\end{center}
Quantum coherence, the physical property underlying fundamental phenomena such as multi-particle interference and entanglement, has emerged as a valuable resource upon which exotic modern technologies are founded. In general, the most prominent adversary of quantum coherence is noise arising from the interaction of the associated dynamical system with its environment. Under certain conditions, however, the existence of noise may drive quantum and classical systems to endure intriguing nontrivial effects. Along these lines, here we demonstrate, both theoretically and experimentally, that when two indistinguishable particles co-propagate through quantum networks affected by noise, the system always evolves into a steady state in which coherences between certain separable states perpetually prevail. Furthermore, we show that the same steady state with surviving quantum coherences is reached irrespectively of the configuration in which the particles are prepared.
\end{abstract}
\maketitle

The influence of random fluctuating environments over the evolution of dynamical systems has been a subject of intensive research since the beginning of modern science \cite{Brown1,Einstein2,McClintock3}. Particularly in quantum physics, environmental noise represents a prominent adversary that precludes the generation, control, and preservation of fundamental properties such as entanglement and quantum correlations \cite{Breuer4,Vega,Aolita,LoFranco}. Conventionally, quantum systems interacting with the environment are termed open quantum systems (OQS), and as such they constitute the most common structures encountered in nature. In this regard, the standard phenomenological approach to describe the evolution of OQS is through the usage of the Born-Markov approximation \cite{Carmichael5}. In such an approach, the system-of-interest is weakly coupled to a large unstructured environment in such a way that the statistical properties of the latter remain unaffected \cite{Hoeppe6,Wiseman7}. Certainly, in the study of OQS one generally restricts to investigate the statistical properties of the system only, that is, to the dynamics of the reduced-OQS \cite{Louisell8,Alicki}. \\

In its simplest configuration, single-particle physics of reduced-OQS subject to Born-Markov premises can be investigated in finite quantum networks in which environmental effects are modeled as pure dephasing \cite{Eisfeld2013}. Mathematically, such networks are described by the stochastic Schr\"odinger equation
\begin{equation}\label{eq:m1}
-\mathrm{i}\frac{d}{dz} \psi_{n}(z)=\beta_{n}(z) \psi_{n}(z) + \sum_{m\ne n}^{N}\cc_{m,n}\psi_{m}(z).
\end{equation}
Here, $\hbar=1$, $\p_{n}$ represents the single-particle wavefunction at site $n$, and $\cc_{m,n}$ are the hopping rates between the $(m,n)$ sites. Moreover, $\ep_{n}(t)$ denotes the fluctuating site energies exhibiting properties of Gauss-Markov processes satisfying the conditions $\langle\ep_{n}(t)=0 \rangle$ and $\langle\ep_{n}(t)\ep_{m}(t')\rangle=\g_{n}\delta_{m,n}\delta(t-t')$, with $\langle...\rangle$ denoting stochastic average and $\g_{n}$ representing the dephasing rates \cite{Eisfeld2013}. The relevance of this model has been highlighted in an interdisciplinary framework of studies ranging from biology \cite{Plenio10,Rebentrost11}, quantum chemistry \cite{Park12}, and electronics \cite{Leon-Montiel13} to photonics \cite{Saikin14,Biggerstaff15} and ultra-cold atoms \cite{Schonleber16}. Yet, despite this model fundamentally describes the behavior of single-particle OQS, it does not show any divergence from wave mechanics \cite{Yu17,Torres18}. Indeed, the richness and complexity of genuine quantum processes are more prominet when a manifold of indistinguishable particles are jointly considered \cite{Sansoni,Matthews19,Walter,Franco}. A crucial fact to emphasize about reduced-OQS is that they involve stochastically fluctuating potentials (site energies) due to the interaction with the environment. Hence, without loss of generality, one can engineer reduced-like OQS in the laboratory by coupling various subsystems each one endowed with fluctuating site energies \cite{Bender20}. That is, the dynamics of true reduced-OQS is effectively reproduced by a set of subsystems in which the fluctuating parameters are implemented physically \cite{Bender20}. 	

In the present work, we investigate theoretically and experimentally reduced Born-Markov OQS within the single- and two-excitation manifolds. In particular, we discus the role of particle indistinguishability in the evolution of identical particles, bosons and fermions, traversing reduced Born-Markov OQS. For our experiments, reduced-like OQS are implemented using waveguide networks inscribed in fused silica glass by means of the femtosecond laser writing technique \cite{Itoh21,Szameit22}. In the context of photonic waveguides, time is mapped to the propagation coordinate $(t\rightarrow z)$, the individual propagation constants, here denoted as $\beta_{n}(z)$ for the $n$-th site, play the role of site energies and the hopping rates result from the evanescent overlap between normal modes supported by adjacent sites \cite{Itoh21,Szameit22}. To produce dephasing effects in the waveguides we induced longitudinal random fluctuations in their refractive indices as displayed in Fig. (1a). These random changes are readily accomplished by varying the inscription velocity of the waveguides at intervals of one centimeter, effectively producing stochastic fluctuations in the site-energies every $\propto$ 33ps.

Throughout this paper, as demonstrative models we consider waveguide trimers involving two relatively strongly-coupled channels both of which interact weakly with a third site, Fig. (1a). The length of the samples for all experiments was 12cm and the propagation constants were taken randomly from a Gaussian distribution with variance $\sigma=3$cm$^{-1}$ $\left(\sigma=2\right.$cm$\left.^{-1}\right)$ for the classical (quantum) experiments, and mean values $\beta_{1}=\beta_{2}=1$cm$^{-1}$ and $\beta_{3}=-1$cm$^{-1}$ for the upper and lower waveguides, respectively. The coupling coefficients were chosen to be $\kappa_{1,2}=\kappa_{2,1}=2$cm$^{-1}$ and $\kappa_{1,3}=\kappa_{2,3}=0.6$cm$^{-1}$. The dephasing rates are estimated using the relation $\g_{m}=\sigma_{m}^{2}\triangle z$ \cite{Laing23}, where $\sigma_{m}$ is the standard deviation used to inscribe the $m$-th waveguide and $\triangle z=1$cm is the correlation length.  

Based on the fact that the optics of single-bosons and single-fermions is analogous to wave mechanics \cite{Yu17}, here we experimentally analyze single-excitation dynamics utilizing laser light (see Fig. 1(b) for a sketch of the setting). For two-boson experiments we use photon pairs generated at a wavelength $\lambda=$815nm using standard type-I spontaneous-parametric-down-conversion source by pumping a BiBO crystal with a $407.5$nm continuous-wave laser diode at $70$mW. We employed commercial V-groove fiber arrays to couple the photons into the chip as well as collecting them at the output facet of individual waveguides. Using high-NA multimode fibers in order to feed the photons to the respective avalanche photodiodes ensure low coupling losses at the output side of the chip. From the data of the photo diodes, the photon probability distribution at the output as well as the inter-channel correlations can be extracted. The two-fermion case is analyzed through numerical integration of the associated master equation.

As reference case for the single-excitation manifold, in Fig. (1c) we depict the experimental intensity evolution of light traversing a noiseless trimer. Evidently, the light propagates in a coherent fashion hopping predominantly among the strongly-coupled channels (upper sites), and at most 10\% of the total energy hops into the farthest (lower) site. In contrast, when the trimers become disturbed by noise, the regular hopping of the wavefunctions is no longer sustained. Instead, the average wave-packets evolve into an incoherent superposition of delocalized light states. These effects are demonstrated experimentally by injecting light into one of the upper sites of an ensemble containing 21 different dynamically disordered trimers. Then, after averaging the intensities over the ensemble we find the pattern displayed in Fig. (1d). These observations unequivocally confirm that single-excitations traversing Markov-Born OQS evolve coherently during certain time, and eventually the system reaches a steady state constituted of a uniform incoherent mixture of states. Remarkably, such homogeneousness in the intensity distribution occurs despite the fact the associated waveguides are inscribed at different separation distances. 

In the presence of noise the proper instrument to describe propagating quantum entities is the density matrix \cite{Wiseman7}. Accordingly, within the single-excitation manifold the dynamics of the average (reduced) density matrix is governed by the master equation (see Appendix)
\begin{equation}\label{eq:m2}
\begin{aligned}
\frac{d}{dz}\left\langle \rho_{n,m}\right\rangle&=\left[\im\left(\beta_{n}-\beta_{m}\right)-\frac{1}{2}\left(\g_{n}+\g_{m}\right)\right]\left\langle \rho_{n,m}\right\rangle+\sqrt{\g_{n}}\sqrt{\g_{m}}\left\langle \rho_{n,m}\right\rangle \delta_{n,m}\\&+\im\sum_{r}\cc_{n,r}\left\langle \rho_{r,m}\right\rangle-\im\sum_{r}\cc_{m,r}\left\langle \rho_{n,r}\right\rangle,
\end{aligned}
\end{equation}
where $\left\langle \rho_{n,m}\right\rangle=\left\langle \p_{n}\p_{m}^{*}\right\rangle$, and $\g_{n,m}$ represents the dephasing rates. In fig. (1) of the Appendix we present a comparison of the numerical diagonal elements $\left\langle \rho_{n,n}\right\rangle=\left\langle \p_{n}\p_{n}^{*}\right\rangle$, computed from Eq. (\ref{eq:m2}), and the experimental intensity distributions shown in Fig. (1d). The excellent agreement between the experimental and numerical results suggest that Eq. (\ref{eq:m2})  can be assumed valid. Moreover, upon inspection of the theoretical off-diagonal elements, $\left\langle \rho_{n,m}\right\rangle$, we identify a gradual loss of coherence. Indeed, these effects occur since the off-diagonal elements $\left\langle \rho_{p,q}\right\rangle$ exhibit a sort of \emph{propagation-constant} $\left(\beta_{p}-\beta_{p}\right)+\im\left(-\g_{p}-\g_{q}\right)/2$, where the imaginary part implies attenuation. This demonstrates that due to attenuation caused by dephasing all the coherences unavoidable decay. For diagonal elements $\left\langle \rho_{p,p}\right\rangle$, such propagation-constants are zero. Fig. (2) depicts the theoretical evolution of $\left\langle \rho_{n,m}\right\rangle$ for different dephasing strengths. Notice in all cases the resulting coherence terms inherently decay demonstrating that noise drives the system to inhabit in stationary states with nullified coherence. 

In stark contrast to single-excitations, when two indistinguishable particles co-propagate in the same stochastic structures, interesting effects occur revealing that some of the corresponding coherence terms resist the impact of noise. To elucidate these effects, we rely on the concept of two-particle probability amplitude $\PP_{p,q}$, which jointly describes two particles at sites (p,q) \cite{Abouraddy25}. In terms of $\PP_{p,q}$ we define the average two-particle density matrix $\left\langle\rho_{(p,q),(p',q')}(z)\right\rangle=\left\langle\PP_{(p,q)}\PP_{(p',q')}^{*}\right\rangle$, whose diagonal elements, $\left\langle\PP_{(p,q)}\PP_{(p,q)}^{*}\right\rangle=\left\langle|\PP_{(p,q)}|^{2}\right\rangle$, give the joint probability density $G{p,q}^{(2)}=\left\langle\rho_{(p,q),(p'q')}\right\rangle$, also termed coincidence rate \cite{Saleh26,Gilead27}.  

In particular, we consider the situation when the system is excited by two indistinguishable particles  $\ket{\PP^{sep}}=\frac{1}{\sqrt{2}}\left(\ket{1_{1},1_{2}}\pm\ket{1_{2},1_{1}}\right)$ (pure two-particle separable states), where the ± signs determine whether the particles are bosons or fermions. The density matrix corresponding to these initial states is shown in Figs. (3 a, j). For bosons we additionally examine the evolution of maximally path-entangled two-particle states, $\ket{\PP^{ent}}=\frac{1}{\sqrt{2}}\left(\ket{1_{1},1_{1}}\pm\ket{1_{2},1_{2}}\right)$, Fig. (3 d). Throughout this work we use the compact notation $\ket{1_{m},1_{n}}$ to represent the state $\ket{1_{m}}\otimes\ket{1_{n}}$, and it represents a state where one particle is at site $m$ and another at $n$. Additionally, states $\propto\left(\ket{1_{m},1_{n}}+\ket{1_{n},1_{m}}\right)$, are symmetrized wavefunctions. At this point, it is worth underlining the off-diagonal terms present in the initial density matrices, Figs. (3 a, j), arise by virtue of the wavefunction simmetrization needed to account for the indistinguishability and exchange statistics of the particles \cite{Feynman28}. Under such excitations, integration of the two-particle master equation (see Appendix), 
\begin{equation}\label{eq:m3}
\begin{aligned}
\frac{d}{dz}\left\langle\rho_{(p,q),(p',q')}(z)\right\rangle=\left[\im \left(\beta_{p}+\beta_{q}-\beta_{p'}-\beta_{q'}\right)-\frac{1}{2}(\g_{p}+\g_{q}+\g_{p'}+\g_{q'})\right]\left\langle\rho_{(p,q),(p',q')}(z)\right\rangle\\
+\left[ \sqrt{\g_{p}\g_{p'}}\delta_{p,p'}+\sqrt{\g_{p}\g_{q'}}\delta_{p,q'}
+\sqrt{\g_{q}\g_{p'}}\delta_{q,p'}+\sqrt{\g_{q}\g_{q'}}\delta_{q,q'}-\sqrt{\g_{p}\g_{q}}\delta_{p,q}-\sqrt{\g_{p'}\g_{q'}}\delta_{p',q'}\right]\left\langle\rho_{(p,q),(p',q')}(z)\right\rangle\\
+\im \sum_{r} \left[\cc_{r,p}\left\langle\rho_{(r,q),(p',q')}(z)\right\rangle+\cc_{r,q}\left\langle\rho_{(p,r),(p',q')}(z)\right\rangle-\cc_{r,p'}\left\langle\rho_{(p,q),(r,q')}(z)\right\rangle-\cc_{r,q'}\left\langle\rho_{(p,q),(p',r)}(z)\right\rangle\right],
\end{aligned}
\end{equation}
renders the average density matrices displayed in Fig. (3). These results clearly show that after a propagation distance of about 12cm, the density matrices for separable and path-entangled bosons become identical, and after 20cm the system reaches its steady state. Once in steady state, a closer inspection of the diagonal elements indicates that both particles bunch into the same site with probability $\left\langle\rho_{(1,1),(1,1)}(z)\right\rangle=\left\langle\rho_{(2,2),(2,2)}(z)\right\rangle=\left\langle\rho_{(3,3),(3,3)}(z)\right\rangle=0.15$, Figs. (3 c, f). The remaining diagonal terms quantify the additional probabilities describing anti-bunching effects, $\left\langle\rho_{(1,2),(1,2)}(z)\right\rangle=\left\langle\rho_{(1,3),(1,3)}(z)\right\rangle=\left\langle\rho_{(2,1),(2,1)}(z)\right\rangle=\left\langle\rho_{(2,3),(2,3)}(z)\right\rangle=\left\langle\rho_{(3,1),(3,1)}(z)\right\rangle=0.09$. \\
Quite interestingly, our theory predicts the existence of quantum coherence in the resultant steady states as indicted by the off-diagonal elements in the density matrices, Figs.~(3 c, f). Indeed, components of the type $\left(\rho_{(p,q),(q,p)},\rho_{(q,p),(p,q)}\right)$, are indistinguishable form each other, as a result they are telltale signs of particle indistinguishability, which is a purely quantum effect. 
However, we must point out that there is an ongoing debate regarding the observability or physical significance of correlations due to symmetrization \cite{Benatti,Reusch}. Indeed, the issue arises because correlations of the type $\left(\rho_{(p,q),(q,p)},\rho_{(q,p),(p,q)}\right)$ represent superpositions of two-particle states where the only difference is the order of the particles. Formally, such coherences do not represent manipulable superpositions, but the presence of such coherences in the steady state imply that the particles retain their capability to interfere in experiments of the Hong-Ou-Mandel type \cite{Killoran30}.

From the two-particle master equation, we see that the density matrix exhibits a complex propagation-constant given by the first two terms on the right-hand-side of Eq.~\eqref{eq:m3}. For the diagonal elements $\left\langle\rho_{(p,q),(p,q)}(z)\right\rangle$ such propagation constant turns out to be zero, and the same occurs for the off-diagonal elements accounting for particle indistinguishability $\left\langle\rho_{(p,q),(q,p)}(z)\right\rangle$. Conversely, for the remaining off-diagonal elements $\left\langle\rho_{(p,q),(p',q')}(z)\right\rangle$ the propagation-constant becomes $\left(-\beta_{p}-\beta_{q}+\beta_{p'}+\beta_{q'}\right)+\im\left(-\g_{p}-\g_{q}-\g_{p'}-\g_{q'}\right)/2$. Owing to the negativity of the imaginary part, we determine that those elements vanish as they are affected by an attenuation factor arising from dephasing. This is the reason for which some coherences decay and some others remain immune to the impact of dephasing. 

Moreover, the steady state can be decomposed into four sub-matrices, $\rho(z)=\rho_{(1,2),(2,1)}^{sep}+\rho_{(1,3),(3,1)}^{sep}+\rho_{(2,3),(3,2)}^{sep}+\rho_{(n,n),(n,n)}^{mix}$, where $\rho_{(n,n),(n,n)}^{mix}\propto \sum_{n=1}^{3}\ket{1_{n},1_{n}}\bra{1_{n},1_{n}}$ represents an incoherent superposition of two-particle probabilities (classically-correlated two-particle state \cite{Abouraddy25}), and $\rho_{(p,q),(q,p)}^{sep}\propto \ket{1_{p},1_{q}}\bra{1_{p},1_{p}} + \ket{1_{p},1_{q}}\bra{1_{q},1_{p}}+H.C.$ is a coherent superpositions of two-particle probability amplitudes. In other words, within the steady state a mixture of both classically-correlated and coherent (indistinguishable) unentangled two-particle states perpetually coexist. Notice, the superposition $\rho_{(1,2),(2,1)}^{sep}+\rho_{(1,3),(3,1)}^{sep}+\rho_{(2,3),(3,2)}^{sep}$ indicate that the particles coherently inhabit in all three sites with the same amplitude.  Nevertheless, such states do not interfere since they form the parts of a stationary state (steady state).

Likewise, for indistinguishable fermion pairs the steady state exhibits some off-diagonal terms Fig. (3 j-l). And the density matrix can be decomposed in a similar fashion as for the boson case indicating the coexistence of quantum superpositions of indistinguishable two-fermion states plus an incoherent mixture of states. Due to the stationarity of the final states (steady states), we infer that they belong to a decoherence-free subspace \cite{Yu17}.

To elucidate the role of particle indistinguishability in the preservation of coherence, we consider the evolution of two-particle states exhibiting classical probabilities, $\rho_{(1,1),(2,2)}^{mix}=\frac{1}{2}\left(\ket{1_{1},1_{1}}\bra{1_{1},1_{1}}+\ket{1_{2},1_{2}}\bra{1_{2},1_{2}}\right)$ and $\rho_{(1,2),(2,1)}^{inc}=\frac{1}{2}\left(\ket{1_{1},1_{2}}\bra{1_{1},1_{2}}+\ket{1_{2},1_{1}}\bra{1_{2},1_{1}}\right)$. Physically, $\rho_{(1,1),(2,2)}^{mix}$ involves two indistinguishable particles entering together into anyone of the upper sites with exactly the same classical probability, i.e., a two-photon states presenting the strongest possible classical correlation \cite{Abouraddy25}. Conversely, $\rho_{(1,2),(2,1)}^{inc}$ represents two distinguishable particles entering separately into the upper sites of the trimer.

Remarkably, for the initial state $\rho_{(1,1),(2,2)}^{mix}$, integration of Eq.~\eqref{eq:m3} renders a steady state which is identical to the ones obtained for separable and path-entangled bosons, Figs.~(3 c, f). Notice, the density matrix for the initial state $\rho_{(1,1),(2,2)}^{mix}$ is not shown here as it is identical to the density matrices shown in Figs.~(3 c, f). In contrast, $\rho_{(1,2),(2,1)}^{inc}$ yields a totally different density matrix which remains incoherent along evolution, Figs.~(3 h, i). From these results we convincingly state that under the influence of dephasing, indistinguishable pairs of particles prepared in any configuration always evolve into the same steady state in which some coherences prevail. In the Appendix we show the resulting two-particle steady states for different dephasing rates, and we demonstrate that the same steady state occurs even in the presence of strong dephasing rates. The extension of the theory to the case of $N$ indistinguishable particles should be straightforward as indicated in the Appendix. \\
For the sake of clarity, we emphasize that single-particle density matrices $\left\langle\rho_{n,m}\right\rangle=\left\langle\p_{n}\p_{m}^{*}\right\rangle$ represent the correlation between single-particle probability amplitudes at sites (n,m). On the other hand, two-particle density matrices $\left\langle\rho_{(p,q),(p',q')}\right\rangle$ describes quantum coherence between the state in which the particles travel in channels $(p,q)$ and $(p',q')$. In both cases, however, the density matrix measures the capability of the associated probability amplitudes to interfere. \\ 
As we mentioned above, the diagonal elements within the density matrices represent the joint particle probability density $G_{p,q}^{(2)}=\left\langle\rho_{(p,q),(p,q)}\right\rangle$. Hence, to prove the validity of the two-particle master equation, Eq. \eqref{eq:m3}, we have performed two-photon intensity correlation measurements for separable, path-entangled, classically correlated, and distinguishable two-photon (two-boson) states using waveguide trimers of 12cm. To prepare indistinguishable separable photon pairs from a SPDC source we additionally apply filters with 3nm bandwidth as shown in Fig. (1b). Path-entangled two-photon states were readily created at the output of an integrated 50:50 directional coupler when simultaneously exciting the two input modes with indistinguishable photons in a separable product state \cite{Lebugle29}. Classically-correlated two-photon states were constructed in a similar fashion with the difference that we induce a delay of $\sim$2ps in one of the output ports of the integrated 50:50 directional coupler such that we have two distinguishable two-photon states (classically-correlated states). Distinguishable two-photon pairs were produced by delaying one of the photons $\sim$2ps with respect to the other before entering the samples (this time without an additional integrated 50:50 directional coupler). We point out that a time delay of $\sim$2ps is sufficient and the distinguishability was verified by the absence of interference in a standard Hong-Ou-Mandel setup.  
The experimental averaged coincidence measurements are depicted in Figs.~(4 e-g) where it is clear that, under the influence of dephasing, initial states involving indistinguishable photons are driven to undergo identical correlation patterns. More specifically, in all three cases separable, path-entangled, and classically correlated two-photon states the measurements reveal the tendency of both photons to bunch into the same site including the farthest weakly-coupled waveguide. Concurrently, photon coincidences occurred with similar probabilities, but less frequently than bunching events as illustrated by the off-diagonal elements in Figs.~(4 e-g). Finally, when exciting the same stochastic networks with distinguishable (incoherent) photons coupled separately into the first and second sites, the correlation patterns were found to exhibit the higher probabilities along the off-diagonal terms, indicating that the state remains incoherent along evolution, Fig.~(4 h). All the averages were taken over an ensemble of 37 samples. In the Appendix we compare the experimental correlations (Fig.~(4)) versus the diagonal elements of the theoretically-computed density matrix displayed in Fig.~(3). Specifically we measured the average fidelity between the experimental $\left\langle G_{p,q}^{(2)-exp}\right\rangle$ and theoretical $\left\langle G_{p,q}^{(2)-th}\right\rangle$ two-particle probability densities $S=\left(\sum_{p,q}\sqrt{\left\langle G_{p,q}^{(2)-exp}\right\rangle\left\langle G_{p,q}^{(2)-th}\right\rangle}\right)^{2}/\sum_{p,q}\left\langle G_{p,q}^{(2)-exp}\right\rangle\sum_{p,q}\left\langle G_{p,q}^{(2)-th}\right\rangle\approx 0.99$. In the present case S is computed over the ensemble average measurements, consequently it is reasonable to obtain a value close to one. \\
In this work, we have investigated, theoretically and experimentally, Born-Markov OQS within the single and two-excitation manifolds. We showed that even when individual particles do not preserve any quantum coherence in the presence of noise, indistinguishable two-particle states preserve, on average, quantum coherence despite the impact of dephasing. More importantly, the prevailing coherence is independent of the actual state launched into the system provided the particles are indistinguishable, as always the same steady state is reached irrespective of the initial configuration.  Our results might provide useful information on the applicability of decoherence to achieve quantum state engineering, quantum simulation, and even universal computation.

\newpage
\begin{figure}[h!]
\centering
\includegraphics[width=14cm]{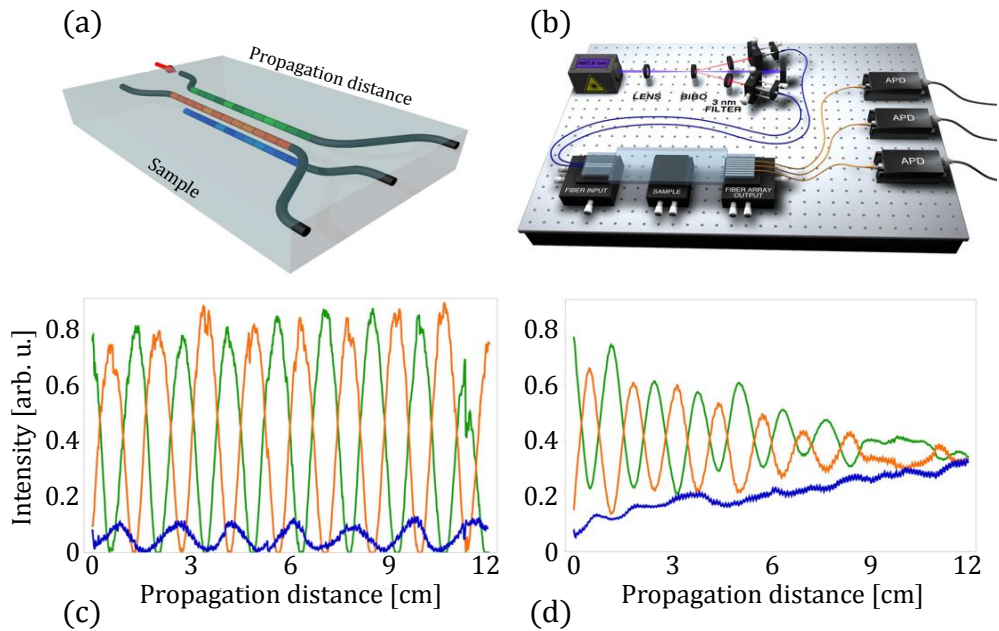}
\caption{(a) Schematic setup of an integrated waveguide trimer simulator for reduced Born-Markov open quantum system. The different colors in the waveguides represent dynamical random changes in the propagation constants, whose effects emulate site-energy fluctuations induced by the environment. (b) Experimental setup employed to carry out experiments within the two-excitation manifold: a two-photon source at a wavelength of 815 nm was implemented by means of spontaneous parametric down-conversion from a pump laser at a wavelength of 407.5 nm. Photons emerging at the output of the device are collected via a fiber array and subsequently fed into avalanche photodiodes (APDs). In the absence of noise all waveguides have the same propagation constants. Consequently, exciting one of the uppers sites with laser light creates the intensity dynamics shown in (c). We observe that light propagates through the system hopping predominantly among the upper waveguides, i.e., the strongly coupled sites. In the presence of dephasing, we observe the emergence of a uniform redistribution of energy among all the waveguides in (d). These experiments unequivocally demonstrate that within the single-excitation manifold dephasing induces a uniform redistribution of energy and as a result the farthest waveguide becomes populated with about 1/3 of the total energy.}
\label{fig:F1}
\end{figure}
\newpage
\begin{figure}[t!]
\centering
\includegraphics[width=14cm]{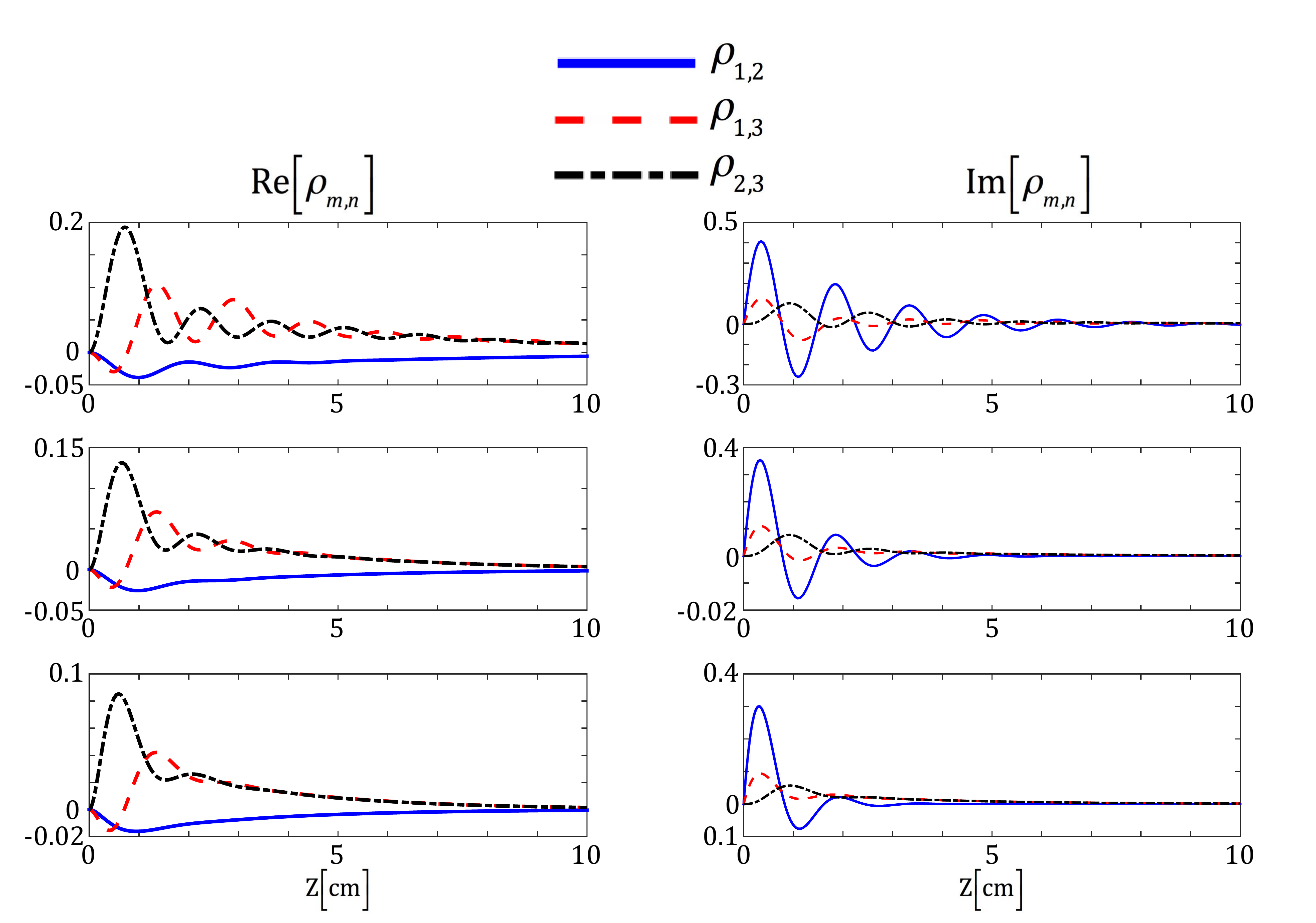}
\caption{Theoretically calculated evolution of reduced density matrices arising in waveguide trimers endowed with dephasing rates (top) $\g=0.3\g_{exp}$, (center) $\g=0.6\g_{exp}$, and (bottom) $\g=\g_{exp}$, where $\g_{exp}=\left(\g_{1},\g_{2},\g_{3}\right)=\g_{exp}=\left(1.7275,1.7435,1.7645\right)$cm$^{-1}$ are the dephasing rates obtained from the experimental parameters used to fabricate the structures utilized in the experiments shown Fig. (1 d). In all cases the coherence terms $\left\langle \rho_{m,n}\right\rangle$ inherently vanish after certain propagation distance. Notice here we show only the coherence terms (off-diagonal terms $\left\langle \rho_{m,n}\right\rangle$, $m\neq n$). The length of the samples was 12cm and the propagation constants were taken from a Gaussian distribution with identical variance $\sigma=3$cm$^{-1}$, and mean values $\beta=1$cm$^{-1}$  and $\beta=-1$cm$^{-1}$ for the upper and lower waveguides, respectively. }
\label{fig:F2}
\end{figure}
\begin{figure}[t!]
\centering
\includegraphics[width=12cm]{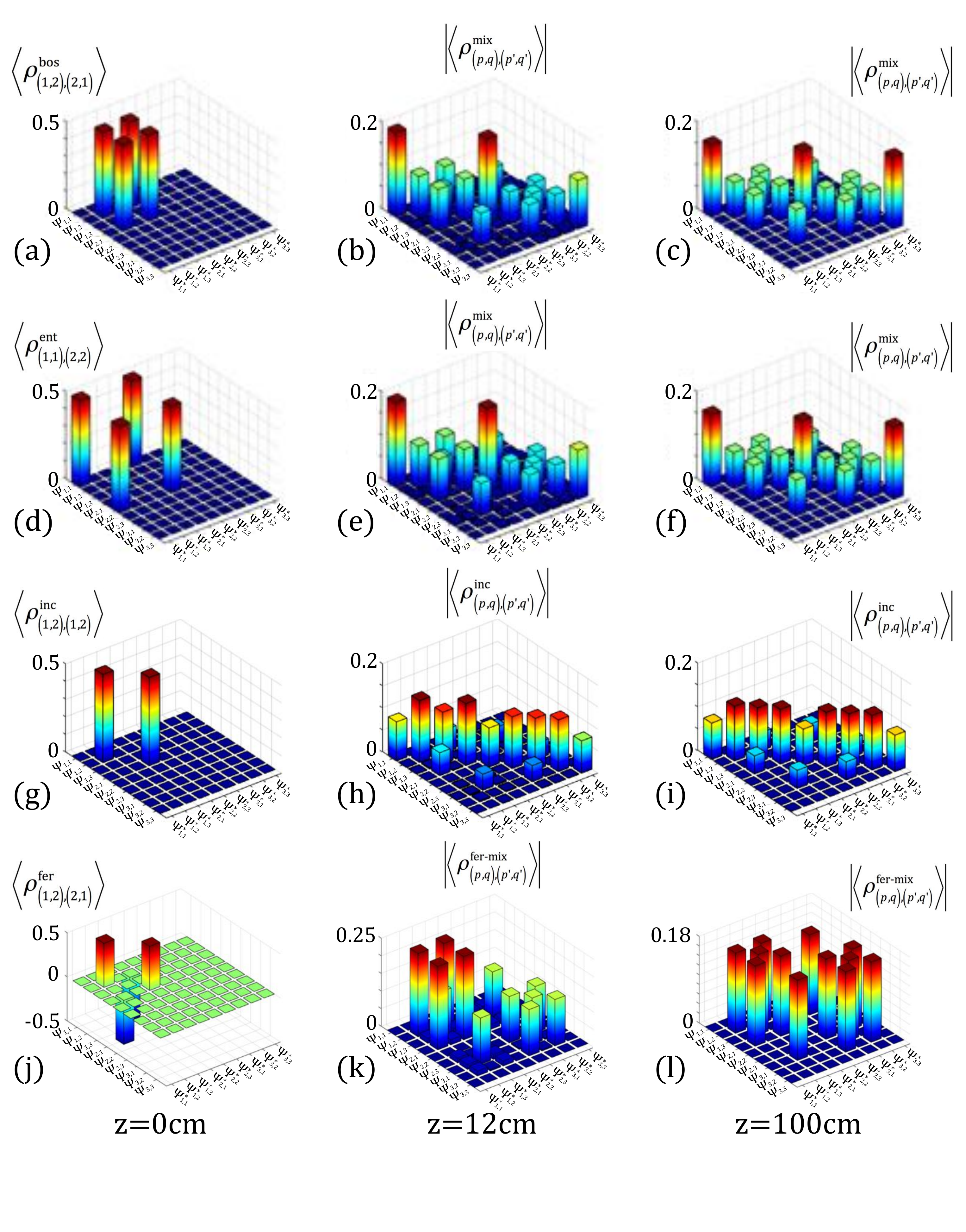}
\caption{Reduced density matrices for separable 
$\ket{\PP^{sep}}=\frac{1}{\sqrt{2}}\left(\ket{1_{1},1_{2}}\pm\ket{1_{2},1_{1}}\right)\rightarrow \rho_{(1,2),(2,1)}=\ket{\PP^{sep}}\bra{\PP^{sep}}$ 
(a), path-entangled 
$\ket{\PP^{ent}}(0)=\frac{1}{\sqrt{2}}\left(\ket{1_{1},1_{1}}\pm\ket{1_{2},1_{2}}\right)\rightarrow \rho_{(1,1),(2,2)}^{ent}(0)=\ket{\PP^{ent}}\bra{\PP^{ent}}$
 (d), and incoherent 
 $\rho_{(1,2),(2,1)}^{inc}(0)=\frac{1}{2}\left(\ket{1_{1},1_{2}}\bra{1_{1},1_{2}}+\ket{1_{2},1_{1}}\bra{1_{2},1_{1}}\right)$ 
 (g) bosons propagating in the noisy trimer shown in Fig.1 (a). The dephasing rates are $\g_{exp}=\left(\g_{1},\g_{2},\g_{3}\right)=\left(1.3012,1.2365,1.293\right)$cm$^{-1}$. We see that at $z=$12cm, separable and path-entangled bosons are described by identical density matrices (b, e). At $z$=100cm the steady state exhibits three main peaks along the diagonal, indicating that photon bunching is the most probable outcome to occur (c, f). In contrast, incoherent photon pairs exhibit completely different density matrices after $z$=12cm (h) and $z$=100cm (i), that is, incoherent superpositions remain incoherent along evolution. Figs. (j-l) depict density matrices for indistinguishable fermion pairs 
$\ket{\PP^{sep}}=\frac{1}{\sqrt{2}}\left(\ket{1_{1},1_{2}}-\ket{1_{2},1_{1}}\right)\rightarrow \rho_{(1,2),(2,1)}=\ket{1_{1},1_{2}}\bra{1_{1},1_{2}}-\ket{1_{1},1_{2}}\bra{1_{2},1_{1}}+H.C.$. From this figure we see a steady state containing off-diagonal entries demonstrating that some coherences survive the impact of dephasing.}
\label{fig:F3}
\end{figure}
\newpage
\begin{figure}[t!]
\centering
\includegraphics[width=14cm]{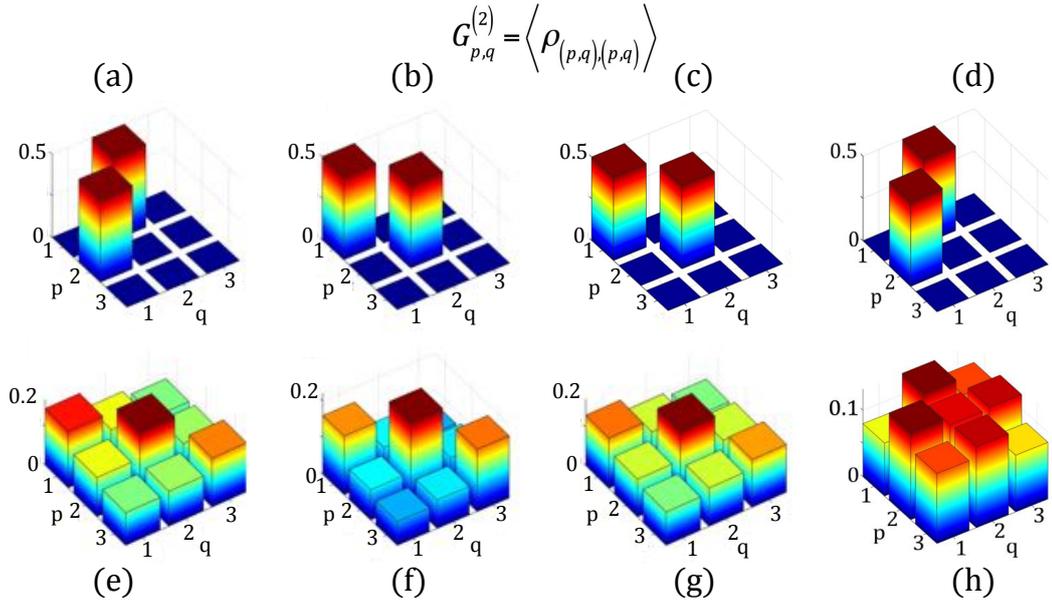}
\caption{Experimental intensity correlation functions, $G_{p,q}^{(2)}=\left\langle\rho_{(p,q),(p,q)}\right\rangle$, for separable (a), path-entangled (b), classically correlated (c), and incoherent photon pairs (d) coupled into the upper sites of the waveguide trimer shown in Fig. (1 a). At $z$=12cm the initial states (a), (b), (c), and (d) exhibit the correlation patterns shown in (e), (f), (g), and (h), respectivelly. Note the correlation matrices, $G_{p,q}^{(2)}=\left\langle\rho_{(p,q),(p,q)}\right\rangle$, have been arranged in a way that the bunching elements, 
$\left(\left\langle\rho_{(1,1),(1,1)}\right\rangle,\left\langle\rho_{(2,2),(2,2)}\right\rangle,\left\langle\rho_{(3,3),(3,3)}\right\rangle\right)$, are shown along the diagonal, while the anti-bunching terms, 
$\left(\left\langle\rho_{(1,2),(1,2)}\right\rangle,\left\langle\rho_{(1,3),(1,3)}\right\rangle,\left\langle\rho_{(2,1),(2,1)}\right\rangle,\left\langle\rho_{(2,3),(2,3)}\right\rangle,\left\langle\rho_{(3,1),(3,1)}\right\rangle,\left\langle\rho_{(3,2),(3,2)}\right\rangle\right)$, are displayed in the off-diagonals entries. Evidently, indistinguishable two-photon states produce similar correlation patterns, while incoherent distinguishable photons exhibit totally different correlations. }
\label{fig:F4}
\end{figure}
\newpage
\appendix*
\section{APPENDIX:}
\setcounter{equation}{0}
In the following, we derive the evolution equations for density matrices arising within the single- and two-excitation manifolds.
\subsection{Single-Excitation Manifold}
We start by considering a stochastic network containing $N$ coupled sites.
In such configurations, the propagation dynamics of single-particle probability amplitudes are governed by the stochastic Schr\"odinger equation \cite{Eisfeld2013}
\begin{equation}\label{eq:1}
-\mathrm{i}\frac{d}{dz} \psi_{n}(z)=\beta_{n}(z) \psi_{n}(z) + \sum_{m\ne n}^{N}\cc_{m,n}\psi_{m}(z).
\end{equation}
Here we have set $\hbar=1$, $\psi_{n}(z)$ is the probability amplitude for a single-particle propagating through site $n$, $\beta_{n}(z)$ are the stochastic site energies which depend on the propagation distance $z$, and $\cc_{m,n}$ represents the coupling coefficients connecting sites $m$ and $n$.
In order to account for environmental effects, we assume random site energies varying according to the functions $\beta_{n}(z)=\beta_{n}+\phi_{n}(z)$, where $\phi_{n}(z)$ describes a stochastic Gaussian process satisfying the conditions
\begin{subequations}
\begin{equation}\label{eq:2a}
 \begin{aligned}
\langle \phi_{n}(z)\rangle&=0,
 \end{aligned}
\end{equation}
\begin{equation}\label{eq:2b}
 \begin{aligned}
\langle \phi_{n}(z) \phi_{m}(z')\rangle&=\gamma_{n}\delta_{m,n}\delta(z-z'),
 \end{aligned}
\end{equation}
\end{subequations}
with $\langle...\rangle$ denoting stochastic average. Note we have assumed the simplest scenario in which the system is affected by white noise, described by Eq.~\eqref{eq:2b}, where $\g_{n}$ denotes the noise intensity, $\delta_{n,m}$ is the Kronecker delta used to indicate that each site energy fluctuates independently from each other, and $\delta(z-z')$ is a Dirac delta describing the Markovian approximation \cite{Jacobs_book}.\\
 Writing Eq.~\eqref{eq:1} in differential form we have
\begin{equation}\label{eq:3}
 \begin{aligned}
d\psi_{n}=\mathrm{i}\beta_{n}\psi_{n}dz+\mathrm{i}\sum_{r}\cc_{n,r}\psi_{r}dz+\mathrm{i}\psi_{n}\phi_{n}(z)dz.
 \end{aligned}
\end{equation}
In turn, by introducing the Wiener increments
\begin{equation}\label{eq:4}
 \begin{aligned}
dW_{n} &= \frac{\phi_{n}(z)}{\sqrt{\gamma_{n}}}dz,\\
\bigl< dW_{n}dW_{m}\bigr>&=\bigl< \frac{\phi_{n}(z)}{\sqrt{\gamma_{n}}}\frac{\phi_{m}(z)}{\sqrt{\gamma_{m}}}\bigr> dz=\frac{\gamma_{n}}{\gamma_{n}}\delta_{n,m}dz=\delta_{n,m}dz,
 \end{aligned}
\end{equation}
we can cast Eq.~\eqref{eq:3} as
\begin{equation}\label{eq:5}
\begin{aligned}
d\p_{n}&=\im \beta_{n}\p_{n}dz+\im\sum_{r}\cc_{n,r}\p_{r}dz+\im\p_{n}\sqrt{\g_{n}}dW_{n}.
\end{aligned}
\end{equation}
We note Eq.~\eqref{eq:5} has the so-called Stratonovich form \cite{vanKampen1981}. In order to compute the differential of the density matrix, $\rho_{m,n}(z)=\p_{n}(z)\p_{m}^{*}(z)$, we can use Ito's product rule $d(\p_{m}\p_{n}^{*})=d(\p_{n})\p_{m}^{*}+\p_{n}d(\p_{m}^{*})+d(\p_{n})d(\p_{m}^{*})$ \cite{vanKampen1981}, which demands $d\p_{n}$ to be written in Ito's form \cite{Eisfeld2013}
\begin{equation}\label{eq:6}
\begin{aligned}
d\p_{n}&=\left(\im \beta_{n}\p_{n}+\im \sum_{r}\cc_{n,r}\p_{r}-\frac{1}{2}\g_{n}\p_{n}\right)dz+\im \sqrt{\g_{n}}\p_{n}dW_{n}.
\end{aligned}
\end{equation}
Hence, using Eq.~\eqref{eq:6} we obtain the expression
\begin{equation}\label{eq:7}
\begin{aligned}
d(\p_{n}\p_{m}^{*})=\left[\im\left(\beta_{n}-\beta_{m}\right)-\frac{1}{2}\left(\g_{n}+\g_{m}\right)\right]\p_{n}\p_{m}^{*}dz+\im\sum_{r}\cc_{n,r}\p_{r}\p_{m}^{*}dz-\im\sum_{r}\cc_{m,r}\p_{n}\p_{r}^{*}dz\\+\im\sqrt{\g_{n}}\p_{n}\p_{m}^{*}dW_{n}-i\sqrt{\g_{m}}\p_{n}\p_{m}^{*}dW_{m}
+\sqrt{\g_{n}}\sqrt{\g_{m}}\p_{n}\p_{m}^{*}dW_{n}dW_{m},	
\end{aligned}
\end{equation}
where we have only considered terms up to first order in $dz$. Finally, by taking the stochastic average of Eq.~\eqref{eq:7} we arrive to the evolution equation for the single-particle density matrix, which is Eq.~(1) presented in our paper
\begin{equation}\label{eq:8}
\begin{aligned}
\frac{d}{dz}\left\langle \rho_{n,m}\right\rangle&=\left[\im\left(\beta_{n}-\beta_{m}\right)-\frac{1}{2}\left(\g_{n}+\g_{m}\right)\right]\left\langle \rho_{n,m}\right\rangle+\sqrt{\g_{n}}\sqrt{\g_{m}}\left\langle \rho_{n,m}\right\rangle \delta_{n,m}\\&+\im\sum_{r}\cc_{n,r}\left\langle \rho_{r,m}\right\rangle-\im\sum_{r}\cc_{m,r}\left\langle \rho_{n,r}\right\rangle.
\end{aligned}
\end{equation}
\subsection{Two-Excitation Manifold}
We now follow a similar procedure to the one described above to derive the evolution equation governing two-particle density matrices in coupled networks affected by dephasing.
To do so, we start by considering pure two-particle probability amplitudes at sites $p$ and $q$ within a network comprising $N$ sites \cite{leija2017}
\begin{equation}\label{eq:9}
\begin{aligned}
\PP_{p,q}(z)=\sum_{m,n}^{N,N}\varphi_{m,n}\left[U_{p,n}(z)U_{q,m}(z)\pm U_{p,m}(z)U_{q,n}(z)\right],
\end{aligned}
\end{equation}
where $\varphi_{m,n}$ is the initial probability amplitude profile $\left(\sum_{n,m}|\varphi_{m,n}|^{2}=1\right)$, and $U_{r,s}(z)$ represents the impulse response of the system, that is, the unitary probability amplitude for a particle traveling into site $r$ when it was initialized at site $s$. Moreover, the sign $+$ and $-$ determine whether the particles are bosons or fermions.\\
From Eq.~\eqref{eq:9} we define the two-particle density matrix $\rho_{(p,q),(p',q')}(z)=\PP_{p,q}(z)\PP_{p',q'}^{*}(z)$ \cite{leija2017}. And using the Ito's product rule we compute the $z$-derivative of the density matrix
\begin{equation}\label{eq:10}
\begin{aligned}
\frac{d}{dz}\left[\rho_{(p,q),(p',q')}\right]=\left[\frac{d}{dz}\PP_{p,q}(z)\right]\PP_{p',q'}^{*}(z)+ \PP_{p,q}(z)\left[\frac{d}{dz}\PP_{p',q'}^{*}(z)\right]+\left[\frac{d}{dz}\PP_{p,q}(z)\right]\left[\frac{d}{dz}\PP_{p',q'}^{*}(z)\right].
\end{aligned}
\end{equation}
To obtain Eq.~\eqref{eq:10} we need the Ito's form for the differential $\left[d\PP_{p,q}(z)\right]$, which is given by
\begin{equation}\label{eq:11}
\begin{aligned}
d\PP_{p,q}(z)=\im \left(\beta_{p}+\beta_{q}\right)dz \PP_{p,q}(z) +\im \sum_{r} \left(\cc_{r,p} \PP_{r,q}(z)+\cc_{r,q} \PP_{p,r}(z)\right)dz -\frac{1}{2}(\g_{p}+\g_{q})\PP_{p,q}(z)dz\\
+\im\left( \sqrt{\g_{p}}dW_{p}+ \sqrt{\g_{q}}dW_{q}\right) \PP_{p,q}(z)-\sqrt{\g_{p}\g_{q}}\PP_{p,q}(z)dW_{p}dW_{q}.
\end{aligned}
\end{equation}
Eq.~\eqref{eq:11} can be easily obtained by taking the derivative of Eq.~\eqref{eq:9} and using the fact that $U_{r,s}(z)$ are single-particle probability amplitudes which obey Eq.~\eqref{eq:5}, namely
\begin{equation}\label{eq:12}
\begin{aligned}
dU_{p,n}&=\im \beta_{p}U_{p,n}dz+\im\sum_{r}\cc_{p,r}U_{r,n}dz+\im U_{p,n}\sqrt{\g_{p}}dW_{p}.
\end{aligned}
\end{equation}
Then, after some algebra we obtain the evolution equation for the average two-particle density matrix
\begin{equation}\label{eq:13}
\begin{aligned}
\frac{d}{dz}\left\langle\rho_{(p,q),(p',q')}(z)\right\rangle=\left[\im \left(\beta_{p}+\beta_{q}-\beta_{p'}-\beta_{q'}\right)-\frac{1}{2}(\g_{p}+\g_{q}+\g_{p'}+\g_{q'})\right]\left\langle\rho_{(p,q),(p',q')}(z)\right\rangle\\
+\left[ \sqrt{\g_{p}\g_{p'}}\delta_{p,p'}+\sqrt{\g_{p}\g_{q'}}\delta_{p,q'}
+\sqrt{\g_{q}\g_{p'}}\delta_{q,p'}+\sqrt{\g_{q}\g_{q'}}\delta_{q,q'}-\sqrt{\g_{p}\g_{q}}\delta_{p,q}-\sqrt{\g_{p'}\g_{q'}}\delta_{p',q'}\right]\left\langle\rho_{(p,q),(p',q')}(z)\right\rangle\\
+\im \sum_{r} \left[\cc_{r,p}\left\langle\rho_{(r,q),(p',q')}(z)\right\rangle+\cc_{r,q}\left\langle\rho_{(p,r),(p',q')}(z)\right\rangle-\cc_{r,p'}\left\langle\rho_{(p,q),(r,q')}(z)\right\rangle-\cc_{r,q'}\left\langle\rho_{(p,q),(p',r)}(z)\right\rangle\right].
\end{aligned}
\end{equation}
The generalization to $N$ indistinguishable particles is straightforward following similar steps as for the two-particle case and introducing the $N$-particle probability amplitude 
\begin{equation}\label{eq:14}
\begin{aligned}
\PP_{p,q,r,...}(z)=\sum_{a,b,c,...}^{N}\varphi_{a,b,c,...}\left[\chi_{a,b,c,...}^{p,q,r,...}+\chi_{a,b,c,...}^{per}+...\right],
\end{aligned}
\end{equation}
where we have defined $\chi_{a,b,c,...}^{p,q,r,...}=U_{p,a}(z)U_{q,b}(z)U_{r,c}(z)...$, with $U_{m,n}$ representing  the probability amplitude for a single-particle at site $n$ when it was launched at channel $m$, and the superscript $per$ means cyclic permutations of superscripts $p$, $q$, $r$, $...$.\\    
In order to integrate Eq.~\eqref{eq:8} and Eq.~\eqref{eq:13} it is necessary to estimate the individual dephasing rates $\g_{m}$ for $m = 1, 2, 3$. This is easily done using the relation $\g_{m}=\sigma_{m}^{2}\triangle z$ \cite{Laing23,Montiel2014}, where $\sigma_{m}$ is the standard deviation of the $m$-th site, $\triangle z$ is the correlation length. To do the simulations shown in Figs. (2) and (3), we have used the $\sigma_{m}$ obtained from the data utilized to inscribe the waveguides
\begin{equation}\label{eq:15}
\begin{split}
\mathbf{\sigma_{exp}}=\left(\sigma_{1}=1.3143cm^{-1}, \sigma_{2}=1.3204cm^{-1}, \sigma_{3}=1.3283cm^{-1}\right), \quad Classical.\\
\mathbf{\sigma_{exp}}=\left(\sigma_{1}=1.1407cm^{-1}, \sigma_{2}=1.112cm^{-1}, \sigma_{3}=1.1371cm^{-1}\right), \quad Quantum.
\end{split}
\end{equation}
In both cases, classical and quantum, $\triangle z=1cm$. Hence, using these $\mathbf{\sigma_{exp}}$  we obtain the individual dephasing rates for numerical integration of Eq.~\eqref{eq:8}  and Eq.~\eqref{eq:13}
 \begin{equation}\label{eq:16a}
 \begin{split}
\mathbf{\g_{exp}}=\left(\g_{1}=1.7275cm^{-1},\g_{2}=1.7435cm^{-1},\g_{3}=1.7645cm^{-1}\right), \quad Classical.\\
\mathbf{\g_{exp}}=\left(\g_{1}=1.3012cm^{-1},\g_{2}=1.2365cm^{-1},\g_{3}=1.2930cm^{-1}\right), \quad Quantum.
\end{split}
\end{equation}

 \subsection{Experimental and Theoretical Comparison}

\begin{figure}[h!]
\centering
\includegraphics[width=14cm]{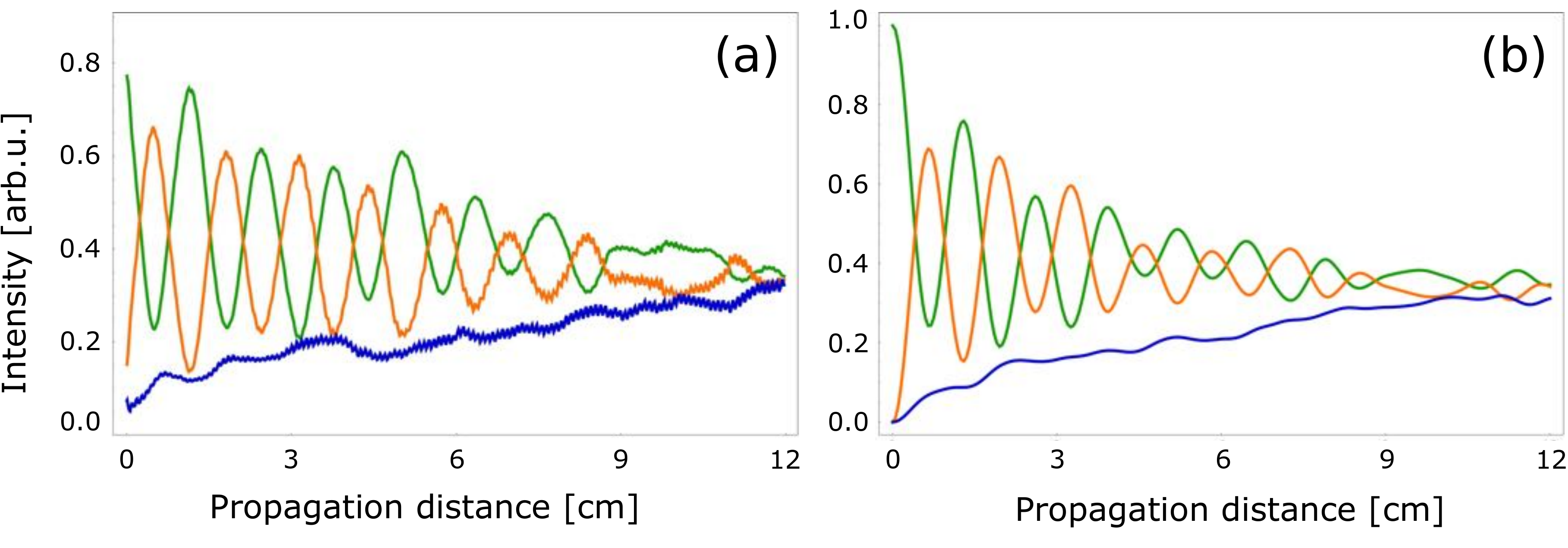}
\caption{Single-excitation dynamics in trimers affected by noise. (a) Experimental intensity distributions. (b) Theoretical probability distribution.}
\label{fig:F11}
\end{figure}

 In this section we provide a comparison between the experimental and theoretical results computed by numerical integration of Eq.~\eqref{eq:8} and Eq.~\eqref{eq:13}. Regarding the single excitation manifold in Fig.~(\ref{fig:F11} a), we show the average experimental intensity distribution recorded after 21 realizations as described in the main text (Fig.~(1 d)). Fig.~(\ref{fig:F11} b) presents  theoretical results obtained from numerical integration of Eq.~\eqref{eq:8}. Notice the excellent agreement between both experimental and theoretical results: the effect of dephasing is to redistribute the energy among all sites, thus producing a uniform distribution after a  propagation distance $z=12$ cm.

\begin{figure}[h!]
\centering
\includegraphics[width=14cm]{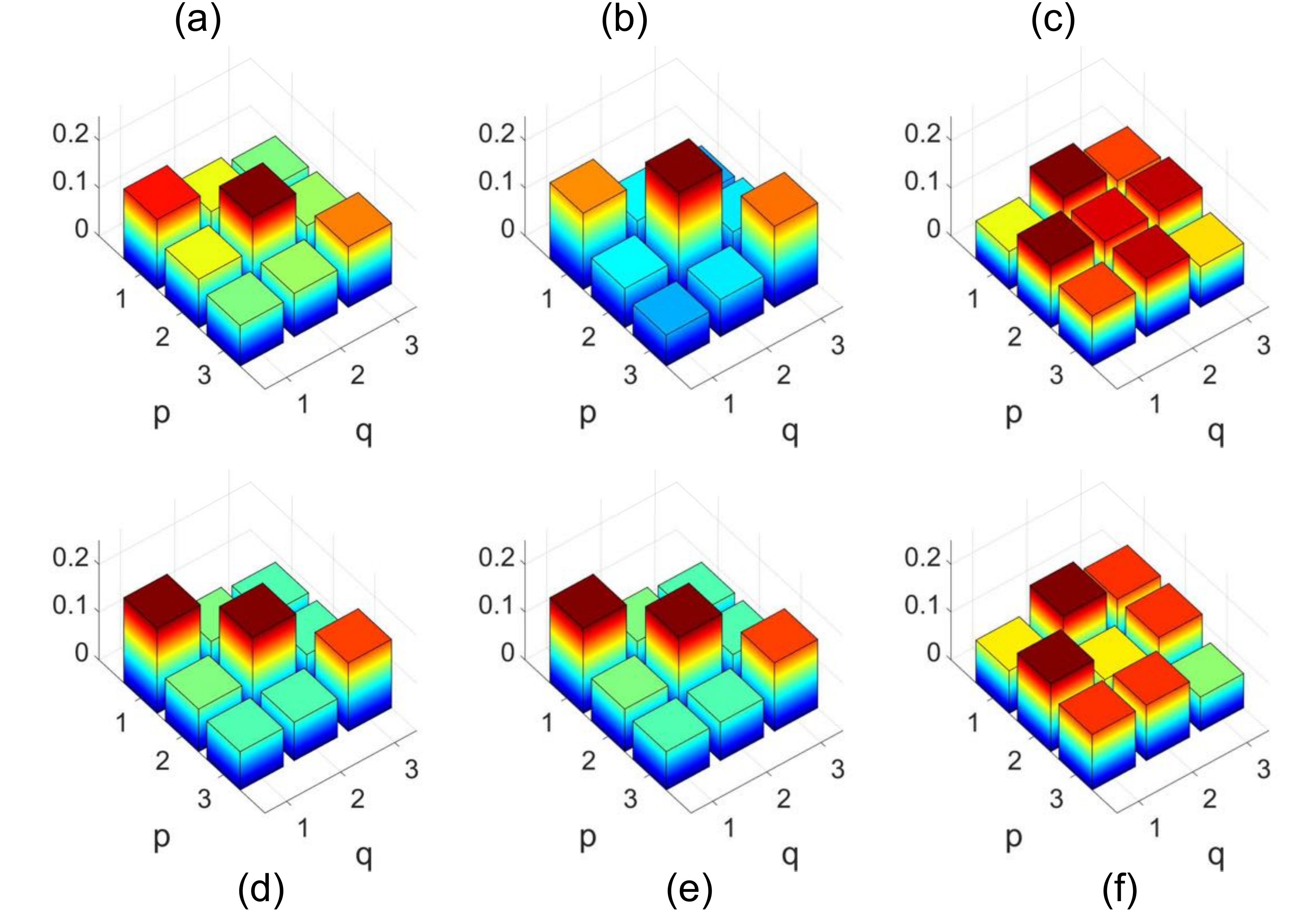}
\caption{Two-particle intensity correlations for separable (a,d), path-entangled (b,e), and incoherent photon pairs (c,e) after $z=12$ cm propagation. Upper row: Experimental data. Lower row: Numerical simulations.}
\label{fig:F22}
\end{figure}

Correspondingly, in Fig.~(\ref{fig:F22}) we show the intensity correlation matrices for separable Figs.~(\ref{fig:F22} a, d), path-entangled Figs.~(\ref{fig:F22} b, e), and distinguishable (incoherent) photon pairs Figs.~(\ref{fig:F22} c, f) after a propagation distance $z=12\text{cm}$. The upper row depicts the experimental coincidence measurements, whereas the lower one shows our theoretical predictions. In order to provide a quantitative comparison between all corresponding cases, we compute the correlation similarity \cite{peruzzo2010,Lebugle29}
\begin{equation}\label{eq:16}
S = \frac{\pare{\sum_{p,q}\sqrt{\langle G_{p,q}^{(2)-\text{exp}}\rangle \langle G_{p,q}^{(2)-\text{th}}\rangle }}^{2}}{\sum_{p,q}\langle G_{p,q}^{(2)-\text{exp}}\rangle\sum_{p,q}\langle G_{p,q}^{(2)-\text{th}}\rangle},
\end{equation}
where $\langle G_{p,q}^{(2)-\text{exp}}\rangle$ and $\langle G_{p,q}^{(2)-\text{th}}\rangle$ stand for the two-particle intensity correlations obtained from the experiments and the theory, respectively. By evaluating Eq. (\ref{eq:16}), one can find that the correlation similarity for all cases is $S\simeq 0.99$, which indicates the high performance of our devices, as well as the validity of our theoretical model.

 \subsection{Impact of strong dephasing on two-photon density matrices}
\begin{figure}[!htb]
\centering
\includegraphics[scale=.5]{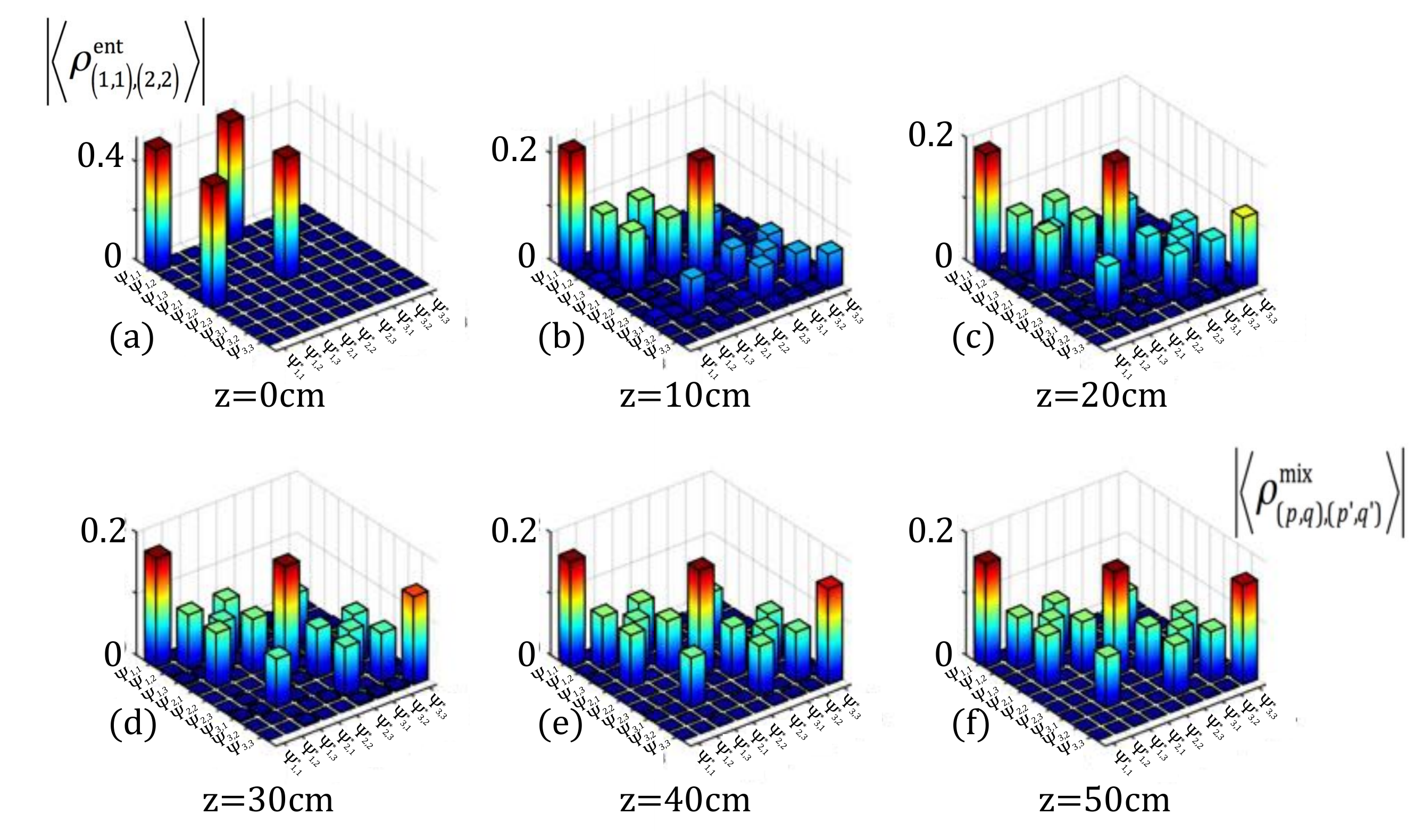}
\caption{Density matrices  (absolute value) for path-entangled two-photon states $\rho^{ent}_{(1,1),(2,2)}(0)$ (a), propagating through waveguide trimers endowed with a dephasing rate of $0.5\g_{exp}$.}
\label{fig:AF2}
\end{figure}

\begin{figure}[!htb]
\centering
\includegraphics[scale=.5]{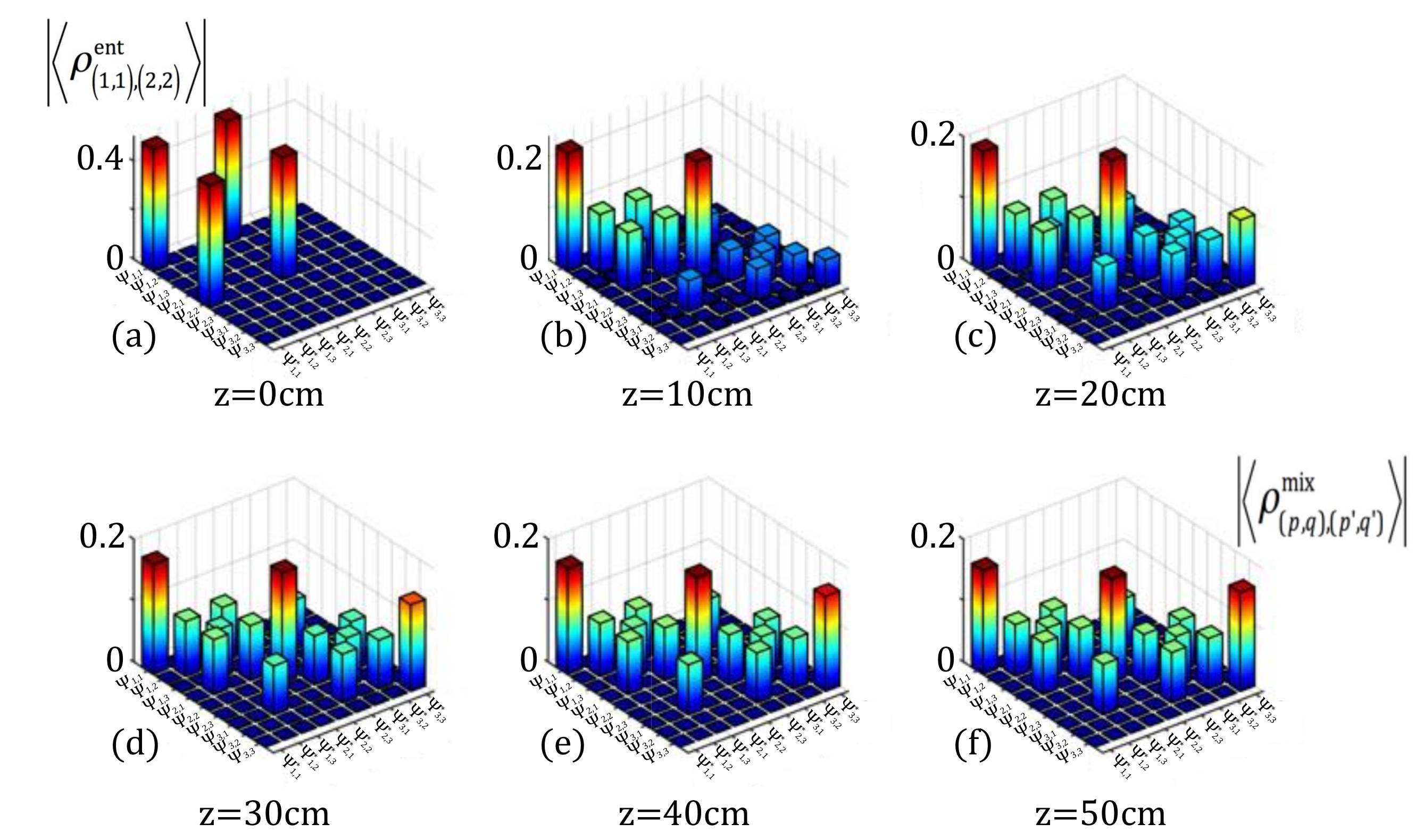}
\caption{Density matrices  (absolute value) for path-entangled two-photon states $\rho^{ent}_{(1,1),(2,2)}(0)$ (a), propagating through waveguide trimers endowed with a dephasing rate of $5\g_{exp}$.}
\label{fig:AF4}
\end{figure}

\begin{figure}[!htb]
\centering
\includegraphics[scale=.5]{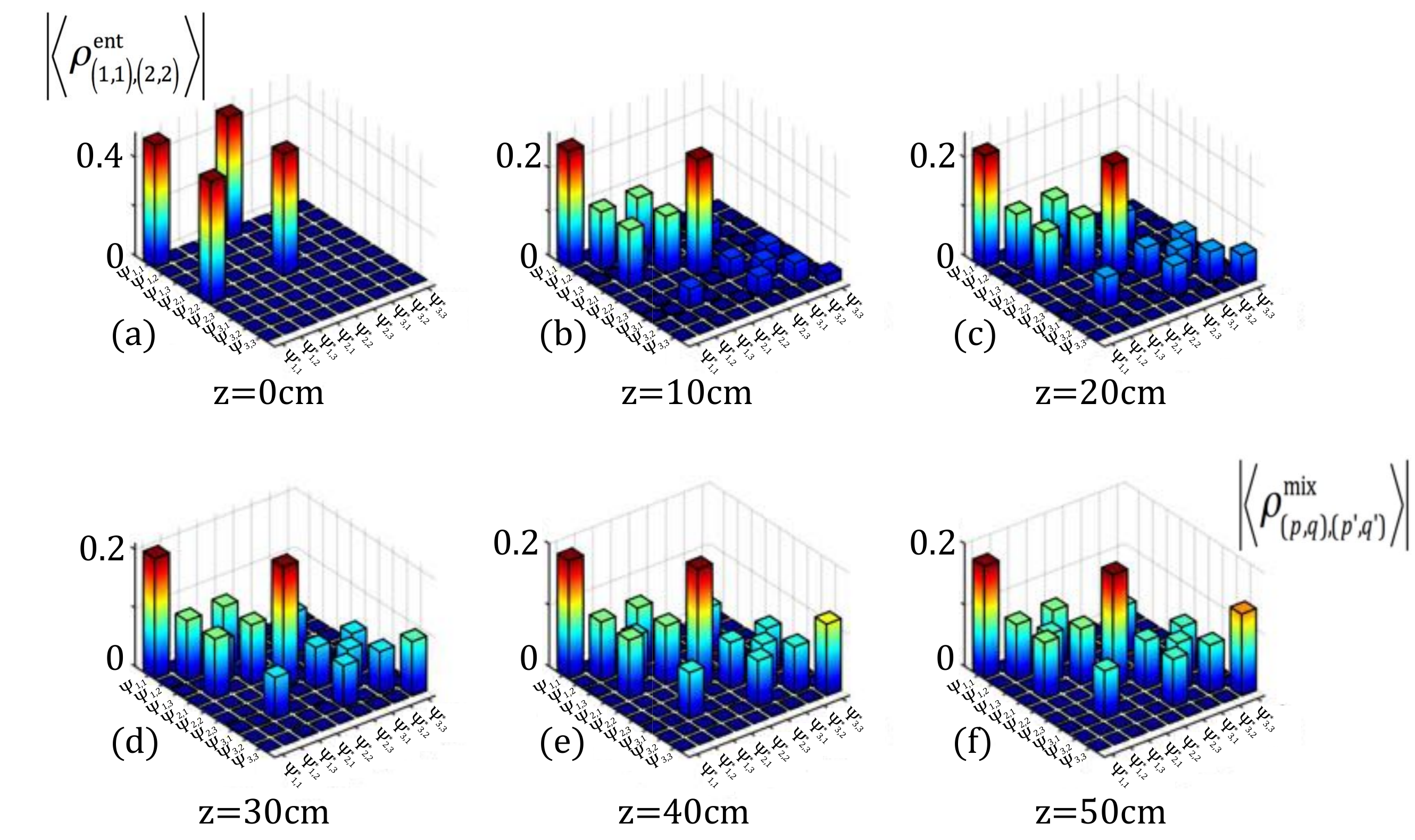}
\caption{Density matrices  (absolute value) for path-entangled two-photon states $\rho^{ent}_{(1,1),(2,2)}(0)$ (a), propagating through waveguide trimers endowed with a dephasing rate of $10\g_{exp}$.}
\label{fig:AF6}
\end{figure}

In order to elucidate the impact of dephasing over path-entangled photon pairs coupled into the upper sites of the waveguide trimer shown in Fig~(1.a) in our paper
\begin{equation}\label{eq:17}
\begin{aligned}
%\rho^{sep}_{(1,2),(2,1)}(0)&=\frac{1}{2}\left(\ket{1_{1},1_{2}}\bra{1_{1},1_{2}}+\ket{1_{1},1_{2}}\bra{1_{2},1_{1}}+\ket{1_{2},1_{1}}\bra{1_{1},1_{2}}+\ket{1_{2},1_{1}}\bra{1_{2},1_{1}}\right),\\
\rho^{ent}_{(1,1),(2,2)}(0)&=\frac{1}{2}\left(\ket{1_{1},1_{1}}\bra{1_{1},1_{1}}+\ket{1_{1},1_{1}}\bra{1_{2},1_{2}}+\ket{1_{2},1_{2}}\bra{1_{1},1_{1}}+\ket{1_{2},1_{2}}\bra{1_{2},1_{2}}\right),
\end{aligned}
\end{equation}
we perform numerical integration of Eq.~\eqref{eq:13} for different dephasing rates. Specifically, we use dephasing rates proportional to $\g_{exp}$,
namely, $\g=0.5\g_{exp}$, $5\g_{exp}$, and $10\g_{exp}$. These values correspond to changing proportionally the variance of the Gaussian distribution utilized to chose the random site energies for the waveguides. For a weak dephasing rate, $\g=0.5\g_{exp}$, integration of Eq.~\eqref{eq:13} renders the density matrices shown in Fig.~(\ref{fig:AF2}). These results indicate that the initial state $\rho^{ent}_{(1,1),(2,2)}(0)$, the steady state emerges at $z\approx40$cm. That is, the system reaches the steady state at twice the distance with respect to the case when $\g=\g_{exp}$ as shown in the main text. For the second case where the dephasing rate is increased to $5\g_{exp}$, Fig.~(\ref{fig:AF4}) indicates that the evolution towards the steady state becomes slower in comparison with the weak dephasing case $(\g=0.5\g_{exp})$. It is important to note however, that such slowing down in the evolution is not substantially noticeable in such dephasing regime. To better appreciate these effects, consider the strong dephasing case, $\g=10\g_{exp}$. In such a scenario, the arising density matrices, shown in Fig.~(\ref{fig:AF6}), clearly illustrate a much slower evolution compared to previous cases. For instance, comparing all density matrices at $z=20$cm, we observe that for $\g=0.5\g_{exp}$ and $\g=5\g_{exp}$, the systems have evolved into very similar states. In contrast, for $\g=10\g_{exp}$ the arising density matrix resembles the one obtained at  $z=10$cm in the previous cases.\\
From these results we can state that in presence of noise, the system will evolve towards the steady state either much slower or much faster depending on the dephasing strength: weak dephasing will drive the system into its steady state faster than strong dephasing will do.
\newpage
%
%\begin{figure}[!htb]
%\centering
%\includegraphics[scale=.5]{Sep1.pdf}
%\caption{Density matrices  (absolute value) for separable two-photon states $\rho^{sep}_{(1,2),(2,1)}(0)$ (a), propagating through waveguide trimers endowed with a dephasing rate of $0.5\g_{exp}$.}
%\label{fig:F1}
%\end{figure}
%
%
%\begin{figure}[!htb]
%\centering
%\includegraphics[scale=.5]{Sep2.pdf}
%\caption{Density matrices  (absolute value) for separable two-photon states $\rho^{sep}_{(1,2),(2,1)}(0)$ (a), propagating through waveguide trimers endowed with a dephasing rate of $5\g_{exp}$.}
%\label{fig:F3}
%\end{figure}
%
%
%\begin{figure}[!htb]
%\centering
%\includegraphics[scale=.5]{Sep3.pdf}
%\caption{Density matrices  (absolute value) for separable two-photon states $\rho^{sep}_{(1,2),(2,1)}(0)$ (a), propagating through waveguide trimers endowed with a dephasing rate of $10\g_{exp}$.}
%\label{fig:F5}
%\end{figure}
%
%

%\section*{References}
\vspace{-1cm}
%\renewcommand{\refname}{\Large{}}

%\vspace{0.75cm}

\end{document}